# Tunneling Spectroscopy Study of Low Temperature Baked Bulk Niobium for Superconducting RF Cavity Applications

I. Curci[1*], G. Jullien[1], L. Maurice[1], F. Eozenou[1], P. Sahuquet[1], and T. Proslier[1]
*1) IRFU, CEA, University Paris-Saclay, F-91191 Gif-sur-Yvette, France*

**ABSTRACT**

Point-contact tunneling (PCT) spectroscopy was used to probe the local superconducting properties of cavity-grade niobium following surface treatments representative of superconducting radio-frequency (SRF) cavities. Electropolished (EP) samples were compared with samples baked at 120 °C for 48 h and subjected to a two-step bake at 75 °C for 2 h followed by 120 °C for 48 h. Although the superconducting gap $\Delta$ is predominantly distributed around the bulk Nb value (~1.5 meV), regions exhibiting strongly suppressed gaps (minigaps) are observed in 15–29% of junctions on EP samples. Their occurrence decreases to 6% after the 120 °C bake and to only 1% after the two-step treatment. Spectra exhibiting minigaps are well described by the Usadel proximity-effect model, consistent with superconducting Nb coupled to normal regions that we attribute to hydride formation. These results provide direct microscopic evidence that low-temperature baking suppresses regions of degraded surface superconductivity in cavity-grade Nb. The pronounced reduction achieved by the two-step bake further supports a link between hydride-related proximity effects and the surface superconducting properties governing SRF cavity performance.

## I. INTRODUCTION

Superconducting radio-frequency (SRF) cavities are key components of modern particle accelerators. Conventional electropolished (EP) niobium cavities routinely achieve quality factors of several $10^{10}$ at operating temperatures of 1.5–2 K and accelerating gradients of 25–30 MV $m^{-1}$ [1–3]. Their maximum accelerating gradient, however, is typically limited by the high-field Q-slope (HFQS), characterized by a rapid degradation of the quality factor at high RF fields [4].

A major advance in mitigating the HFQS was the introduction of the so-called mild bake [5], a low-temperature heat treatment typically performed at 100–140 °C for 12–48 h. This treatment enables EP cavities to reach accelerating gradients of approximately 35–40 MV $m^{-1}$. More recently, a modified two-step bake, incorporating an additional annealing step at 75 °C for 2–4 h before the conventional 120 °C treatment, has produced record accelerating gradients approaching 50 MV $m^{-1}$ [6,7], corresponding to peak surface magnetic fields ~ 215 mT close to the theoretical maximal superheating field ($H_{SH}$) of Nb: $H_{SH} = 1.2H_C = 240$ mT, where $H_C$ is the thermodynamical critical field of Nb ~ 200 mT [8].

Despite its remarkable effectiveness, the microscopic origin of the baking effect remains under debate. Several mechanisms have been proposed to account for the HFQS and its suppression by low-temperature baking, including interstitial oxygen [9], localized hot spots [3,10], dislocations [3], and magnetic impurities [11]. Increasing evidence points towards an important interplay between hydrogen, lattice defects, and oxygen in the near-surface region. In this picture, low-temperature baking modifies the defect structure and promotes oxygen diffusion, creating efficient hydrogen-trapping sites within the first tens of nanometers below the surface and thereby suppressing hydride precipitation [12].

Hydrogen-related degradation in Nb SRF cavities has long been recognized through the phenomenon of Q-disease [13], which results from hydride precipitation during slow cooldown through approximately 75–130 K. Q-disease can largely be avoided through high-temperature hydrogen degassing and sufficiently rapid cooldown. Nevertheless, nanoscale hydrides have been observed even under cooling conditions that suppress conventional Q-disease [12]. Such precipitates may locally degrade superconductivity and thereby limit the maximum accelerating field. Notably, low-temperature baking has been shown to substantially reduce the formation of these nanohydrides [12], suggesting a possible microscopic connection between hydride suppression and improved high-field RF performance.

Because RF currents are confined to within a few magnetic penetration depths of the cavity surface ($\lambda \approx 40$ nm for Nb in the clean limit), cavity performance is particularly sensitive to the superconducting properties of the near-surface region. Understanding how surface processing and low-temperature annealing modify local quantities such as the superconducting energy gap, Δ, and critical temperature, $T_c$, as well as non-ideal signatures arising from proximity effects or inelastic scattering, is therefore essential for identifying the microscopic mechanisms that limit SRF performance. Tunneling spectroscopy provides direct access to the quasiparticle density of states (DOS) and is thus particularly well suited to probing such local modifications of superconductivity.

*Contact author: ivana.curci@cea.fr

Previous tunneling spectroscopy studies have proven useful in establishing a link between local tunneling spectra and the RF performance of SRF cavities. In particular, scanning tunneling microscopy (STM) and spectroscopy (STS) studies were performed to investigate the effect of nitrogen doping on the surface electronic and chemical structures of cutouts from superconducting Nb radio-frequency cavities [14] on ≤ 1 μm lateral scales. However, STM/STS requires a clean sample surface, often prepared in situ, and require the native oxide layer removal. In this work, we use point-contact tunneling (PCT) spectroscopy [15], a complementary tunneling-based approach that probes the superconducting DOS while preserving the native surface oxide intact, potentially providing a more representative picture of the surface state actually present in an SRF cavity. PCT further enables spatially resolved measurements of the superconducting DOS on hundreds of microns lateral scales, providing a sensitive probe of heterogeneous or locally suppressed superconductivity that may not be apparent from smaller scale measurements.

Here, we use PCT spectroscopy to compare the surface superconducting DOS of cavity-grade Nb coupons subjected to processing conditions representative of those used for SRF cavities. Measurements are performed over ~100-µm-scale regions on EP samples, samples subjected to the conventional 120 °C bake, and samples receiving the two-step 75 °C/120 °C treatment. By comparing the distributions of local superconducting properties across these treatments, we investigate the microscopic changes induced by low-temperature baking and their possible connection to the suppression of the HFQS. In particular, we examine whether local spectroscopic signatures can provide insight into the enhanced effectiveness of the two-step bake and its ability to extend the accessible accelerating field of Nb SRF cavities.

## II. EXPERIMENTAL DETAILS

In this study, we report RF measurements of two 1.3 GHz bulk niobium cavities subjected to distinct EP and post thermal treatments in high vacuum (HV). Following EP, one cavity received a bake at 120 °C for 48 h; The other cavity was treated with a two-step bake at 75 °C for 2 h followed by 120 °C for 48 h. The base pressure was 1–2 $10^{-7}$ mbar and during the thermal treatments the vacuum reached 6 $10^{-7}$ mbar. Subsequent to these surface treatments, the cavities received standard high-pressure rinsing (HPR) and were assembled for cryogenic testing in an ISO5 clean room environment at CEA Saclay. The cryogenic testing was performed in the synergium vertical test facility at CEA where the cavities were submerged in a dewar of liquid helium then cooled through pumping to temperatures in the range 1.35-1.7 K.

We characterized the surface superconducting properties of cavity grade bulk niobium coupons processed under conditions identical to those of the corresponding cavities by means of PCT as described in [15]. In our setup, the junctions were formed by approaching the sample surface with an Au tip.

The PCT data were analyzed using the expression for the differential tunneling conductance

$$\frac{dI}{dV} \propto \int N_S(E) \left[-\frac{\partial f(E+eV)}{\partial(eV)}\right] dE \qquad (1)$$

where $I$ is the current flowing through the SIN junction under a difference of potential $V$, $N_S(E)$ is the superconducting DOS, and $f(E)$ is the Fermi function.

Two models were considered to describe $N_S(E)$. The first is the widely known Dynes formula [16],

$$N_S(E) = N_0 \mathrm{Re}\left[\frac{E+i\Gamma}{\sqrt{(E+i\Gamma)^2-\Delta^2}}\right] \qquad (2)$$

where $N_0$ is the DOS at the Fermi surface in the normal state, Δ is the superconducting gap and Γ is the phenomenological quasiparticle lifetime broadening parameter.

*Contact author: ivana.curci@cea.fr

To explicitly account for proximity-induced modifications of the superconducting DOS, the spectra were also analyzed within the framework of the Usadel equations [17], considering a diffusive normal-metal layer (N) proximity coupled to a bulk superconductor (S). The resulting DOS is controlled by fours parameters: the Dynes broadening parameter $\Gamma$, the BCS deep gap $\Delta_D$ in S, and two dimensionless parameters $\alpha$ and $\beta$, which are determined by the N-layer thickness and the N-S interface transparency:

$$\alpha = \frac{d}{\xi_S}\frac{N_N}{N_S},\ \ \beta = \frac{4e}{\hbar}R_B N_N\,\Delta_D d. \tag{3}$$

Here $\xi_S$ is the bulk coherence length, $d$ is the N-layer thickness, $N_N$ and $N_S$ are the DOS at the Fermi surface in the normal state of the N-layer and S-layer, respectively, and $R_B$ is the contact resistance of the N-S interface.

In the Usadel theory, the gap induced in the normal metal layer, often referred to as the minigap $\Delta$ [17], is given by the equation:

$$\Delta = \frac{1}{\Delta/\Delta_D}\left(\frac{1-\Delta/\Delta_D}{1+\Delta/\Delta_D}\right)^{1/2} \tag{4}$$

The minigap $\Delta$ constitutes the effective superconducting gap measured by PCT and is approximately given by half the energy separation between the quasiparticle peaks. The deep gap $\Delta_D$ in the superconductor underneath the normal metal layer is inferred from the fits but not directly probed by tunneling spectroscopy.

## III. RESULTS

### A. RF tests

In Fig. 1a are presented the results of RF tests carried out at CEA on 1.3 GHz niobium cavities, measured at temperatures ranging from 1.35 K to 1.7 K. Fig. 1b shows the corresponding surface resistance $R_s$ as a function of the inverse temperature 1/T, measured at accelerating field of $E_{acc}$=1 MV/m.

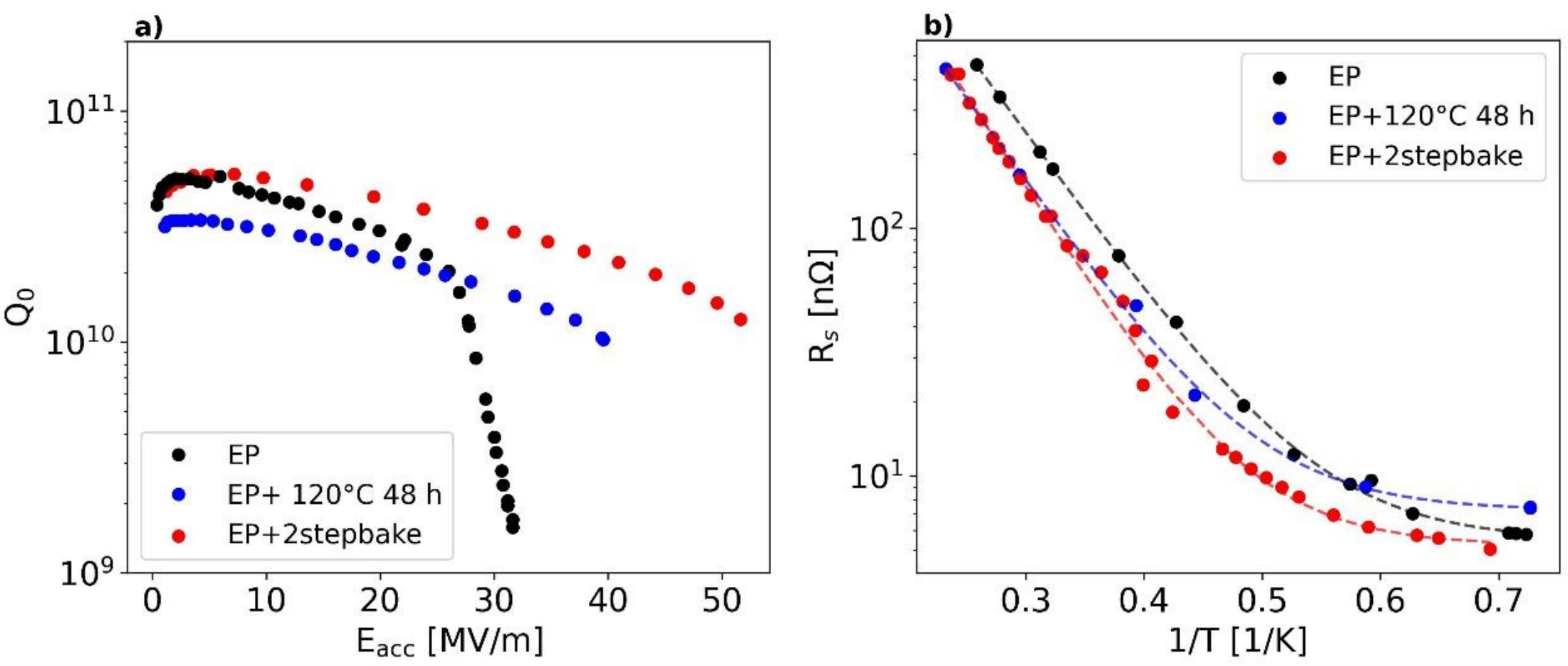

FIG 1. a) $Q_0$ versus $E_{acc}$ results of two 1.3 GHz niobium cavities measured at temperatures in the range 1.35 K-1.5 K. (b) Surface resistance $R_s$ measured as a function of temperature at an accelerating field of 1MV/m (4.3 mT). The dashed lines correspond to fits with Eq. (5).

The experimental data in Fig. 1b were fitted using the expression

$$\mathrm{R}_s = A\exp\left[-\frac{\Delta}{k_B\,\mathrm{T}}\right] + \mathrm{R}_{\mathrm{res}}\ . \tag{5}$$

Eq. (5) describes the surface resistance $R_s$ of a BCS superconductor under an RF field. The first term represents the thermally activated quasiparticle contribution, which depends exponentially on the superconducting energy gap $\Delta$ and the temperature T. $A$ is a pre-exponential factor and $k_B$ is the Boltzmann constant. The second term is the residual resistance $R_{res}$ which is not predicted by BCS theory, but is always observed in experiments. There are several possible candidates that might be related to $R_{res}$, such as trapped magnetic flux, dislocations, and inelastic scattering processes such as magnetic impurities [18–20]. Table I presents a summary of the parameters extracted from the RF tests. The onset of the HQFS $E_{onset}$ and maximum accelerating gradients $E_{MAX}$ were extracted from Fig. 1a. The superconducting energy gap Δ and residual resistance $R_{res}$ were obtained from fits to the data in Fig. 1b using Eq. (5). The extracted parameters are consistent with values commonly reported in the literature [4–5, 21–25].

*Contact author: ivana.curci@cea.fr

TABLE I. Summary of RF performance parameters for the cavities shown in Fig. 1.

| Treatment | $E_{onset}$ [MV/m] | $E_{MAX}$ [MV/m] | Δ [meV] | $R_{res}$ [nΩ] | Literature values | | |
|---|---|---|---|---|---|---|---|
| | | | | | $E_{onset}$ [MV/m] | $E_{MAX}$ [MV/m] | $R_{res}$ [nΩ] |
| EP | 25 | 31 | 1.33(5) | 5.4(2) | 20-25 | 25-30 | 4-7 |
| EP+ mild bake | - | 40 | 1.37(3) | 7.2(2) | - | 35-42 | 6-9 |
| EP + 2stepbake | - | 50 | 1.49(3) | 5.2(3) | - | 48-52 | 4-6 |

## B. TUNNELING RESULTS

In this section, we investigate cavity-grade Nb coupons (grain sizes of 10–100 µm) subjected to EP and low-temperature baking treatments. The main results of this section are summarized in Table II. In Fig. 2, we present selected conductance spectra from a niobium sample that underwent 200 µm of EP, measured before and after the two-step baking treatment. In both cases, most tunnel junctions exhibit typical Nb behavior that can be accurately described by Dynes model; however, some spectra show deviations from this behavior.

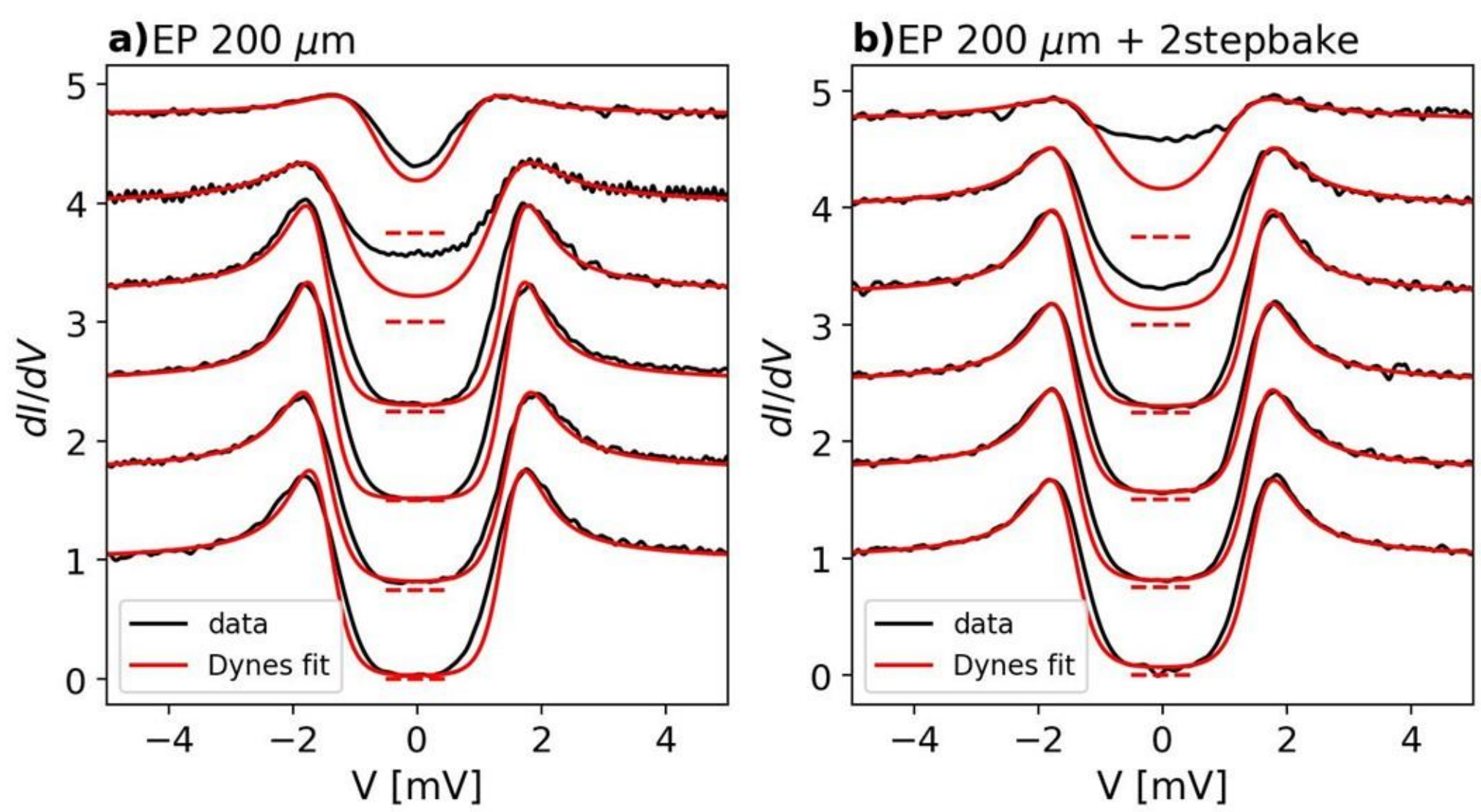

FIG 2: Selected tunneling conductance curves and the corresponding fits from a Nb EP 200 µm sample before a) and after b) two-step bake. The curves have been vertically shifted for clarity. The red dashed lines indicate the zero-conductance level.

For each measurement, approximately 100 junctions were measured over an area of about 100 µm × 100 µm at T= 1.8 K, with a step size of 10 µm, as shown in Fig. 3. It is to be mentioned that the scanned areas are not exactly the same in each measurement, since it goes beyond the optical alignment capabilities of our equipment. We also performed scans over a larger area, as shown in Fig. 4, in order to better assess the homogeneity of the sample over a larger number of grains. Since our samples are fine grain Nb with grain size of ~10-100 µm, we chose a step size of 50 µm for the larger scans.

In Fig. 3, we present the statistics and maps of Δ and Γ/Δ obtained from the scans of 100 µm x 100 µm. In both cases (before and after the two-step bake), the gap distributions peak around the ideal Nb gap value (1.4 to 1.6 meV [26]), but also display a contribution of minigaps (i.e., gaps smaller than the bulk Nb gap, typically below

*Contact author: ivana.curci@cea.fr

1.3 meV), that cannot be fitted with the Dynes model, such as the one illustrated in Fig. 2 and Fig. 6a and that accounts for 32% and 14% of the junctions before and after annealing, respectively. From the larger scan shown in Fig. 4, we observe minigaps in 23% and 18% of the junctions before and after annealing, respectively. In both scales studied, the two-step baking reduces the number of minigap areas. In Fig. 3, we often see that the regions with reduced gap are adjacent, suggesting that the 10 µm step size is comparable to the size of the defects. In order to further investigate these reduced gap area size, we conducted a scan with smaller step size through one of the defects (Fig. 5), from which we estimate a size of approximately ~15 µm.

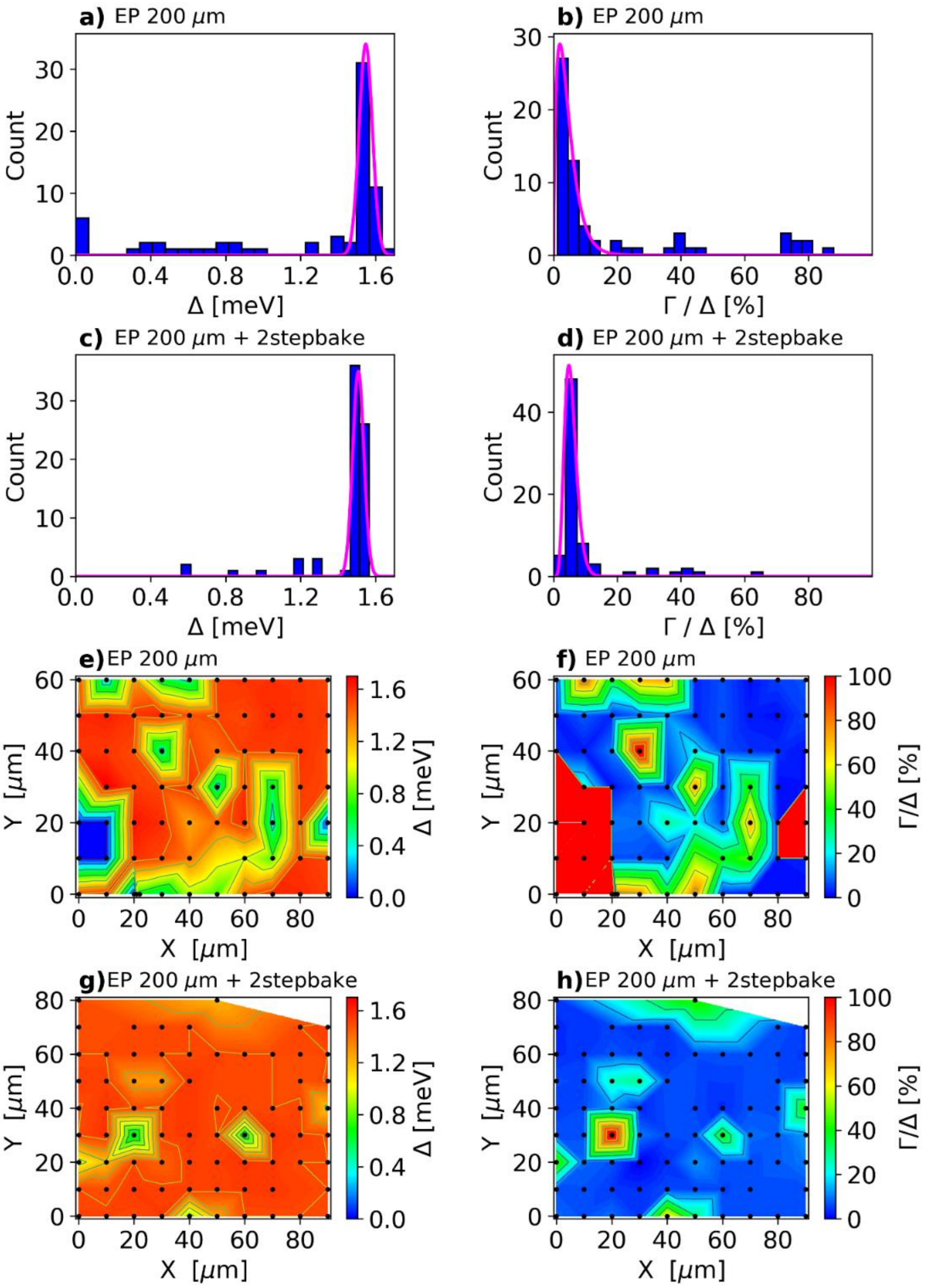

FIG 3: Δ and Γ/Δ statistics and maps obtained from a Nb EP 200 µm sample before and after two-step bake over areas of 100 µm x 100 µm. In magenta, gaussian (a-c) and gamma (b-d) fits. The black dots correspond to the locations of the tunnel junctions.

*Contact author: ivana.curci@cea.fr

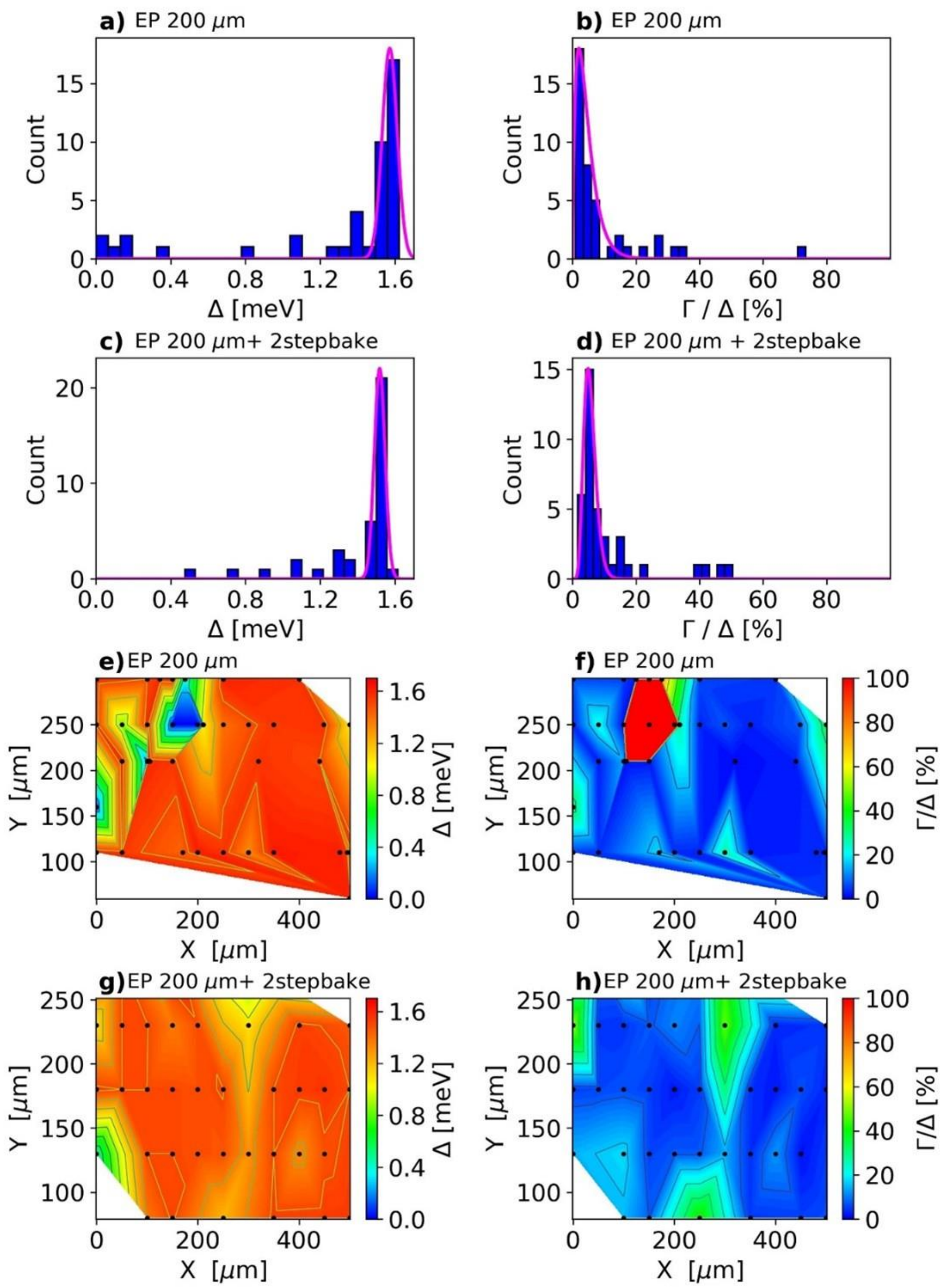

FIG 4: Δ and Γ/Δ statistics and maps obtained from a Nb EP 200 µm sample before and after two-step bake over areas of 500 µm x 300 µm. In magenta, gaussian (a-c) and gamma (b-d) fits. The black dots correspond to the locations of the tunnel junctions.

*Contact author: ivana.curci@cea.fr

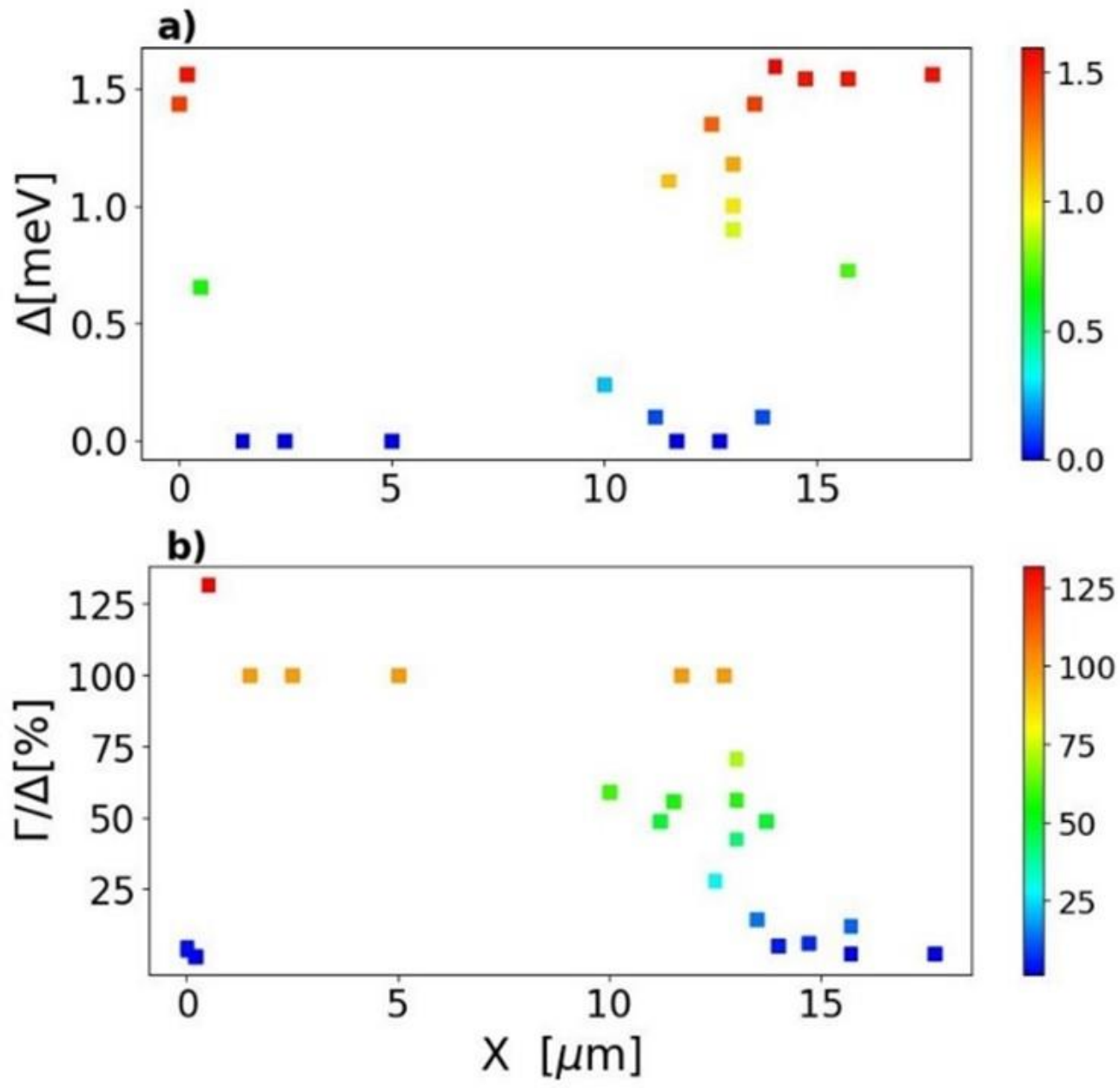

FIG 5: Δ and Γ/Δ parameters extracted from Dynes fits of conductance spectra obtained through a scan along the x-axis on the Nb EP 200 μm sample. The color scale indicates the magnitude of the values.

A typical spectrum with a reduced gap value is represented in Fig. 6a. The spectra is better fitted with the Usadel rather than the Dynes model. The temperature dependence of this junction and the corresponding fits are shown in Fig. 6b; the underlying $\Delta_D$ follows typical BCS behavior expected for ideal Nb with a ratio $2\Delta/k_BT_c \sim 3.7$, similar to the one presented in Fig. 7c. In contrast, the minigap Δ exhibits a significantly reduced gap of 0.86 meV at T=1.8 K, yet superconductivity vanishes at ~9.4 K. This corresponds to $2\Delta/k_BT_c \sim 2.1$, significantly lower than the bulk Nb accepted value. While several mechanisms can be responsible for suppressing the gap, to our knowledge only a proximity effect can account for this specific temperature dependence. This allows us to conclude that in the large majority of reduced gap regions, the gap suppression is due to a proximity effect. However, it is important to note that not all of the minigaps observed can be described by the Usadel proximity model: 75% of the minigaps are well fitted by the Usadel model, while for 11% of them in the 0.5–1 meV range a Dynes fit is already sufficient to describe the data, suggesting that the reduced gap may arise from increased scattering. The remaining 14% cannot be adequately described by either model. The occurrence of spectra that cannot be described by either model has also been reported in STM/STS studies of Nb surfaces [14], highlighting the need for a better understanding of the underlying materials science and surface properties of Nb.

Hydrides are likely candidates to cause the proximity effect, since the cooling conditions during the PCT measurements favor their formation; niobium hydrides form within a temperature range of 75–130 K [13] and during our PCT measurements the samples are held at 85 K for about 2-3 hours using a nitrogen precooling stage before being further cooled to 4 K with helium.

*Contact author: ivana.curci@cea.fr

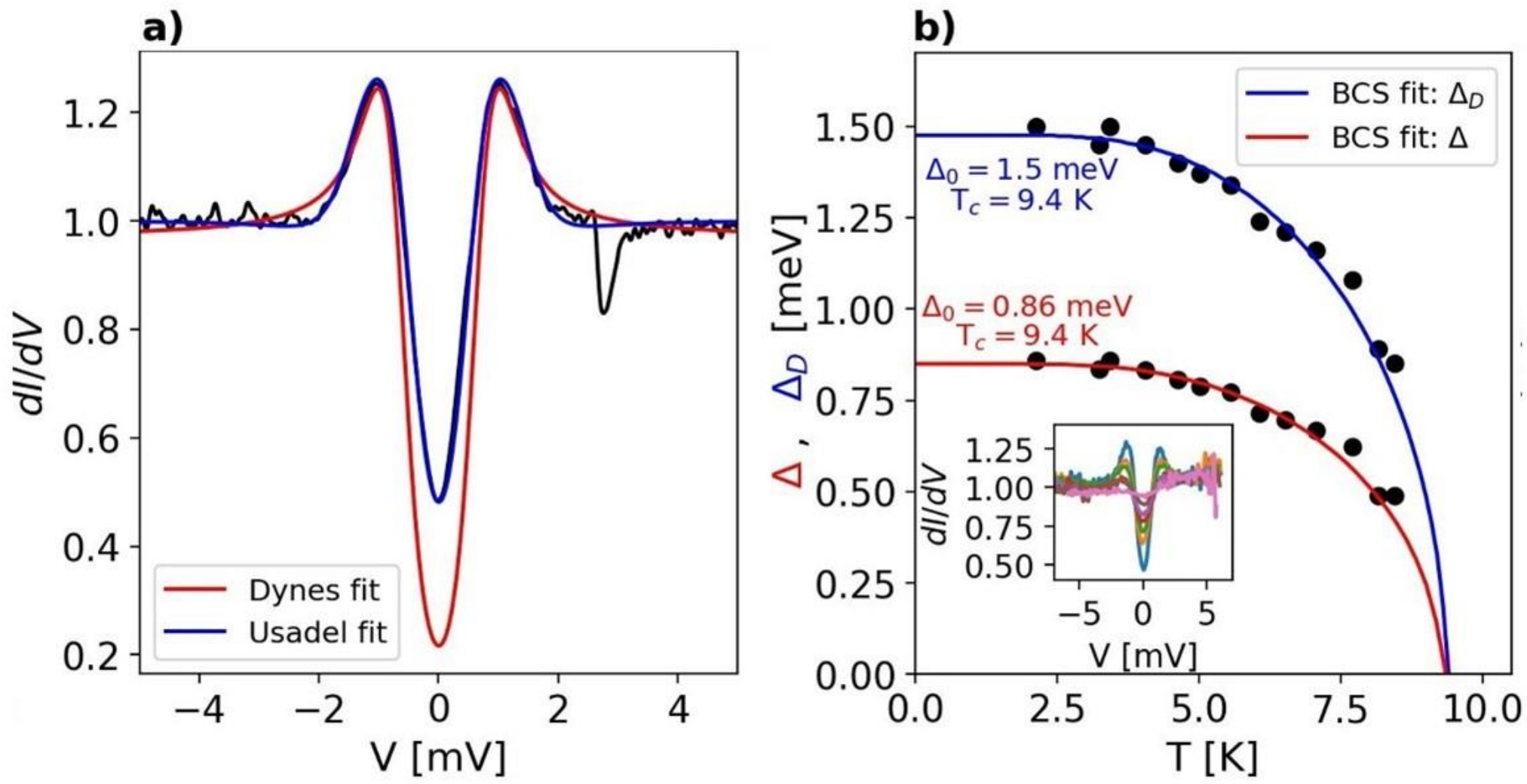

FIG 6: a) Selected junction at 1.8 K with Dynes and Usadel fits for Nb EP 200 µm sample. b) Temperature dependence of $\Delta$ and $\Delta_D$ extracted from Usadel fits.

It is surprising that a proximity effect is still observed after performing the two-step bake. This disagreement with the hypothesis that the two-step bake suppresses hydride precipitation could be due to the fact that we used a slow cooldown protocol. Therefore, if the hydrogen content in the sample is sufficiently high, the conditions could be analogous to those encountered in the so-called Q-disease scenario [13]. This seems plausible, as the size of the defects—on the order of 10 microns—matches the 1–10 µm defect size reported in Q-disease studies [3]. Under these conditions, it has been shown that this phenomenon is non-reversible: even after returning to room temperature, the damage caused by hydride formation leaves permanent traces on the niobium surface. This could explain why we still observe minigaps even after the two-step baking.

In order to rule out any influence of the intermediate warm-up to room temperature on the two-step baking effect, we conducted a second experiment. This time, we measured two different samples: both were treated with EP to remove 100 µm, and one received an additional two-step baking. In Fig. 7, we present selected conductance spectra from both samples. The EP 100 µm sample exhibits spectra similar to those observed in the EP 200 µm one. In contrast, the two-step bake sample displays highly homogeneous junctions, all showing ideal Nb behavior.

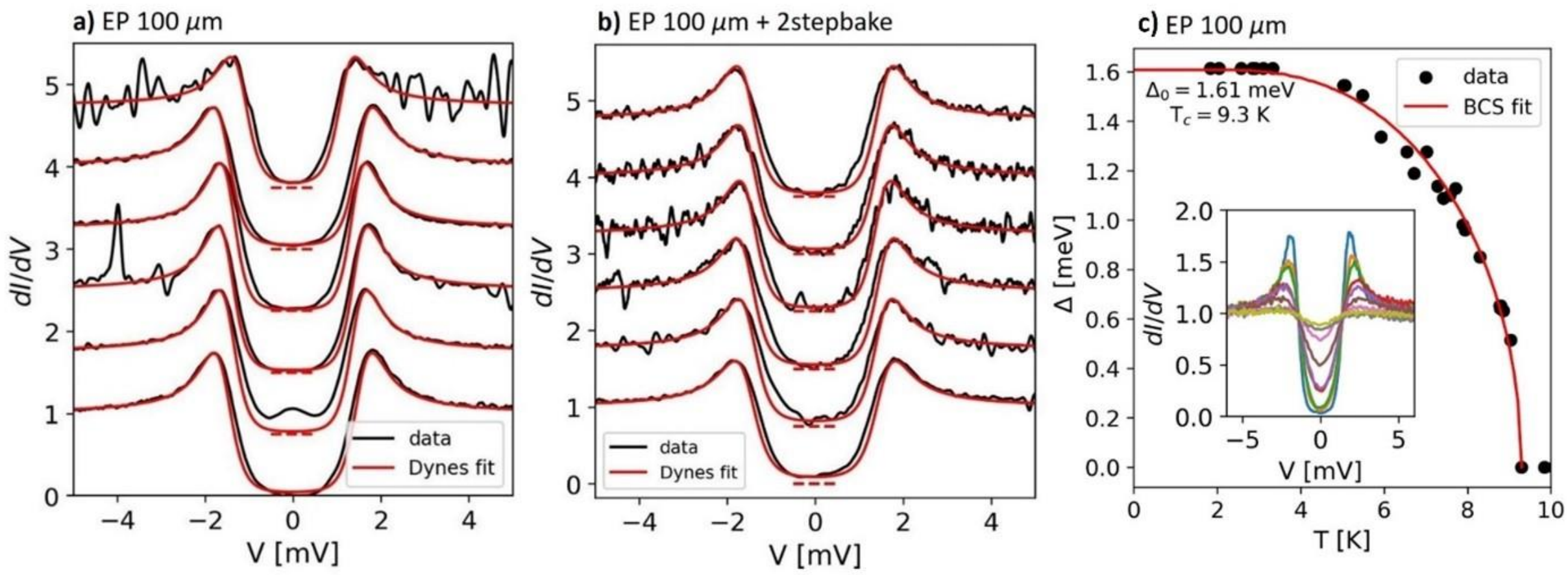

FIG 7: Selected tunneling conductance curves and the corresponding fits from a a) Nb EP 100 µm and b) Nb EP 100 µm + two-step bake samples. The curves have been vertically shifted for clarity. The red dashed lines indicate the zero-conductance level. c) Temperature dependence of $\Delta$ extracted from one measured junction of Nb EP 100 µm sample.

In Fig. 8 we present the statistics and maps of Dynes parameters extracted from fits of conductance spectra measured over an area of 100 µm x 100 µm at T=1.8 K. For the EP 100 µm sample, we still observe the presence of minigaps, although in a less proportion (14%) compared to the EP 200 µm sample (29%). This indicates that, when comparing samples, we must carefully consider that different EP samples may exhibit varying amounts of minigaps, as longer EP will promote more hydrogen diffusion into the Nb. Nonetheless, it is clear that the two-step baked sample is very homogeneous, and that the minigap values have disappeared. We observed a similar mean value ratio, Γ/Δ ∼ 7%, for both two samples, although the two-step baked sample exhibits a smaller dispersion.

*Contact author: ivana.curci@cea.fr

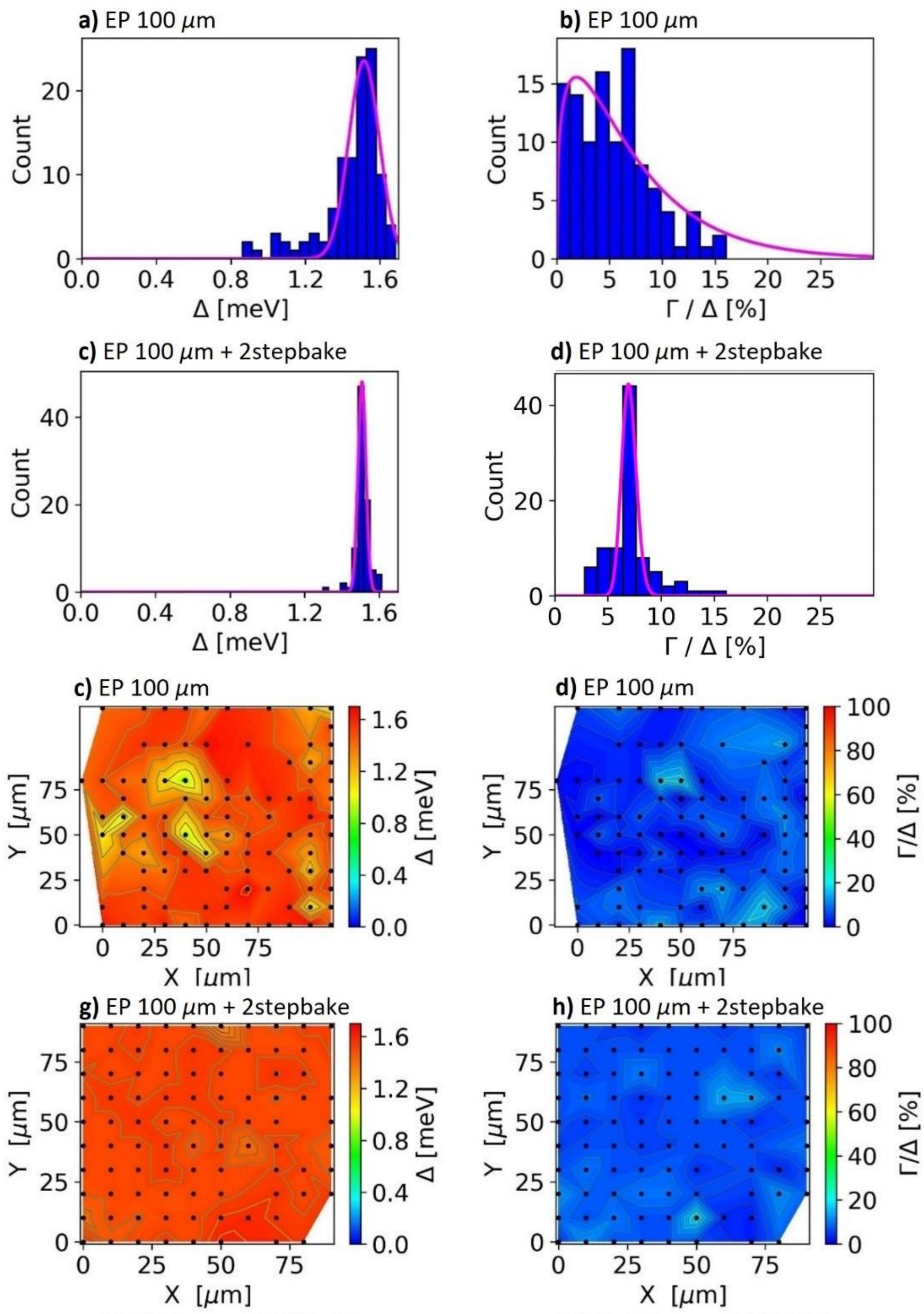

FIG 8: Δ and Γ/Δ statistics and maps obtained from Nb EP 100 µm and Nb EP 100 µm + two-step bake samples. In magenta, gaussian (a-c) and gamma (b-d) fits. The black dots correspond to the locations of the tunnel junctions.

*Contact author: ivana.curci@cea.fr

To test the reproducibility of the two-step bake process we measured a third sample. It is worth mentioning that, in the history of this sample, after removing 100 µm by EP, an annealing treatment was performed at 800°C for 3 hours. To reset the surface, we carried out an EP process to remove the first 10 µm, and subsequently applied the two-step baking treatment. In Fig. 9 we present selected conductance spectra, in which we observe junctions with ideal Nb bulk DOS. The statistics and maps of the parameters extracted from Dynes fits are summarized in Fig.10. The statistic for Δ is similar to those of the two-step baked sample of Fig. 8. However, the ratio Γ/Δ is smaller, with a mean value of ~ 2%, compared to Γ/Δ ~7% in the previous two-step bake sample. Whether this difference in Γ/Δ between the two samples arises from the 800 °C bake or simply reflects sample-to-sample variability requires further investigation.

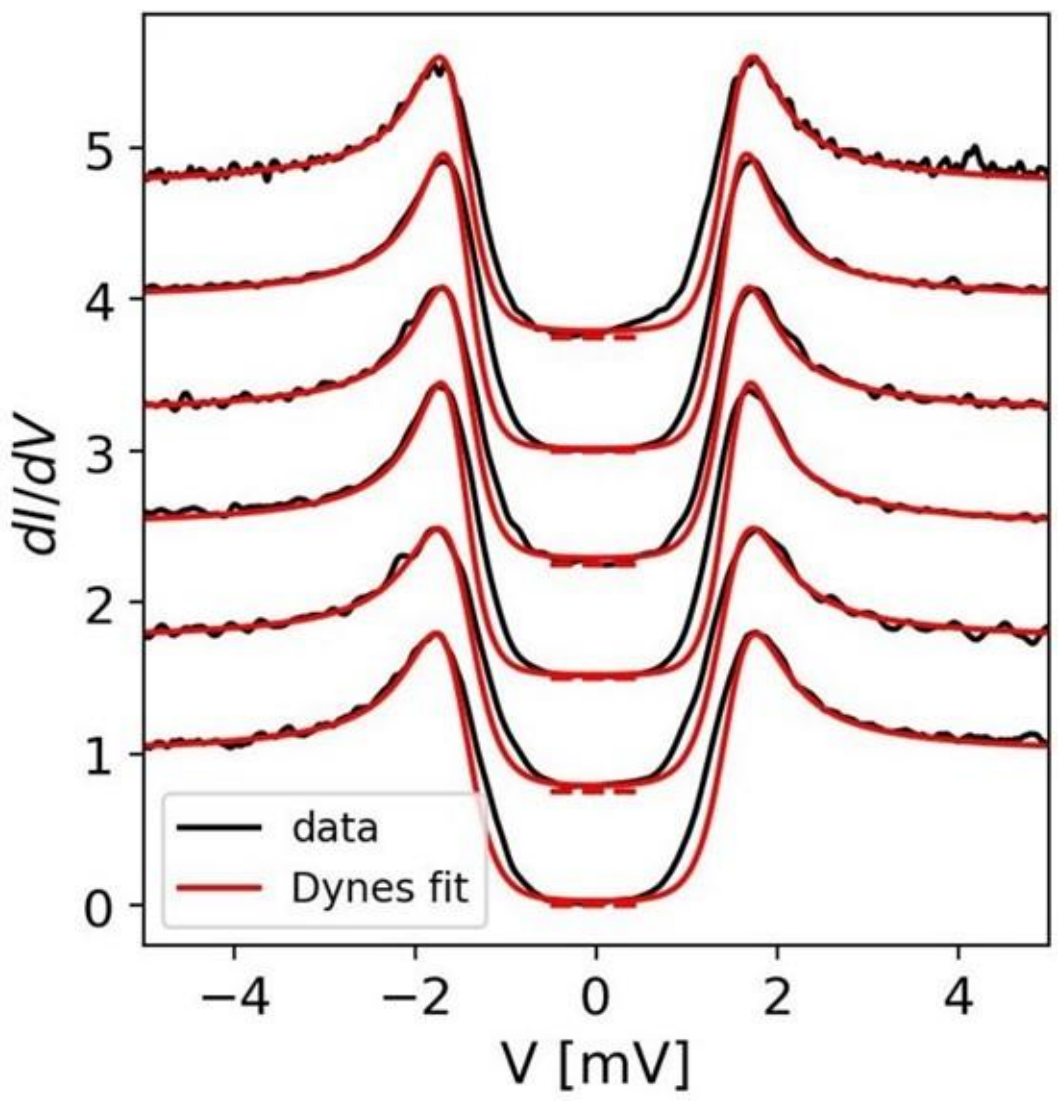

FIG 9: Selected tunneling conductance curves and the corresponding fits from a Nb EP 100 µm + 800°C 3h + EP 10 µm + two-step bake sample. The curves have been vertically shifted for clarity. The red dashed lines indicate the zero-conductance level.

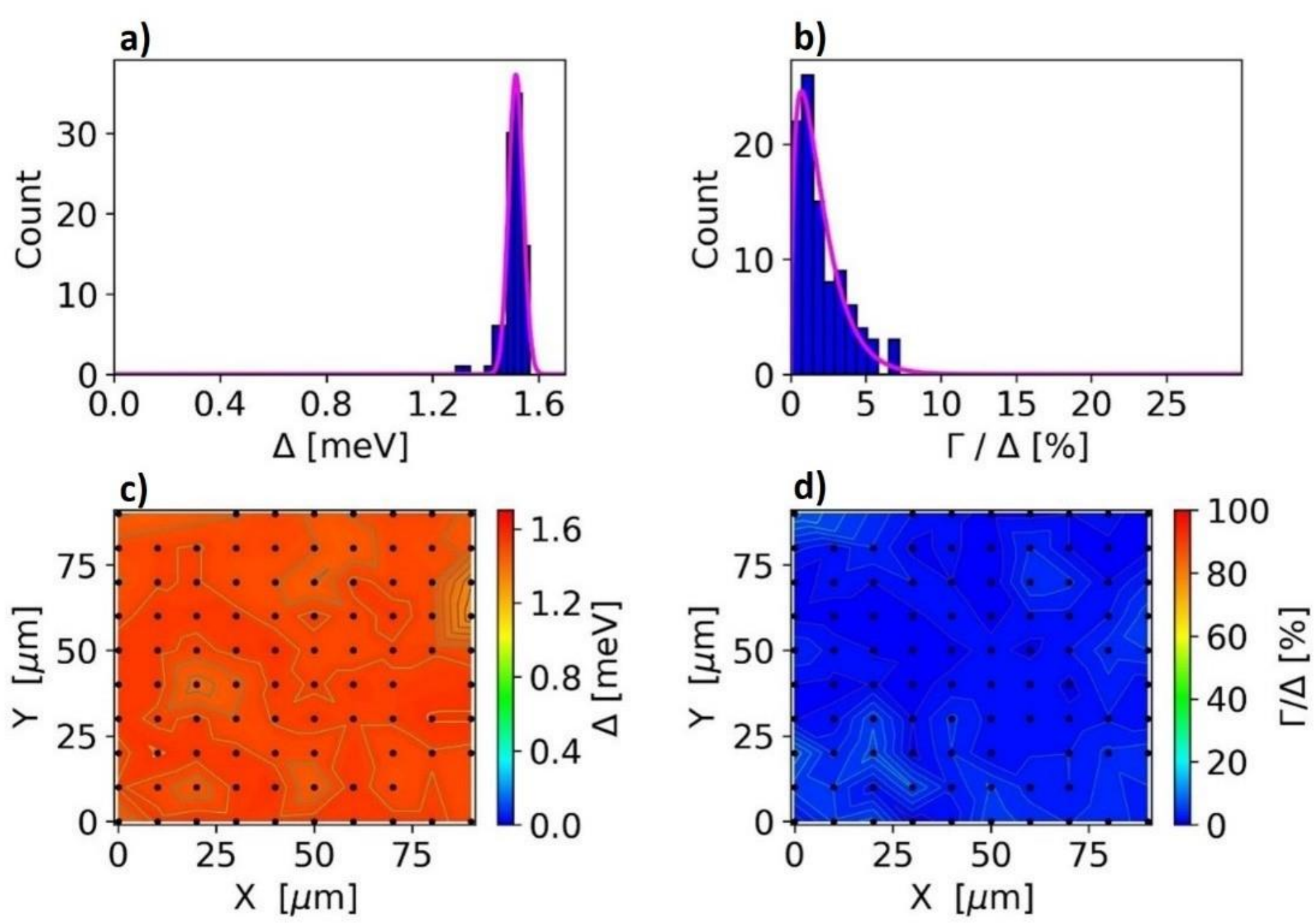

FIG 10: Δ and Γ/Δ statistics and maps obtained from Nb EP 100 µm + 800°C 3h + EP 10 µm + two-step bake sample. In magenta, gaussian (a) and gamma (b) fits. The black dots correspond to the locations of the tunnel junctions.

*Contact author: ivana.curci@cea.fr

In order to compare the effect of the 2step bake with the standard mild baking at 120°C for 48 hrs treatment, we also measured a Nb sample that received 100 µm of EP followed by the standard mild bake at 120°C for 48 h. In Fig. 11 we present selected conductance of this sample. We observe mainly junctions with ideal Nb bulk DOS, similar to the two-step bake samples and the statistics and maps of the parameters (Fig. 12) exhibits similar distributions compare to the two-step bake samples, with the only notable difference that more, ~ 6%, of the junctions in the Nb EP 100 µm + 120 °C 48 h sample display minigap regions.

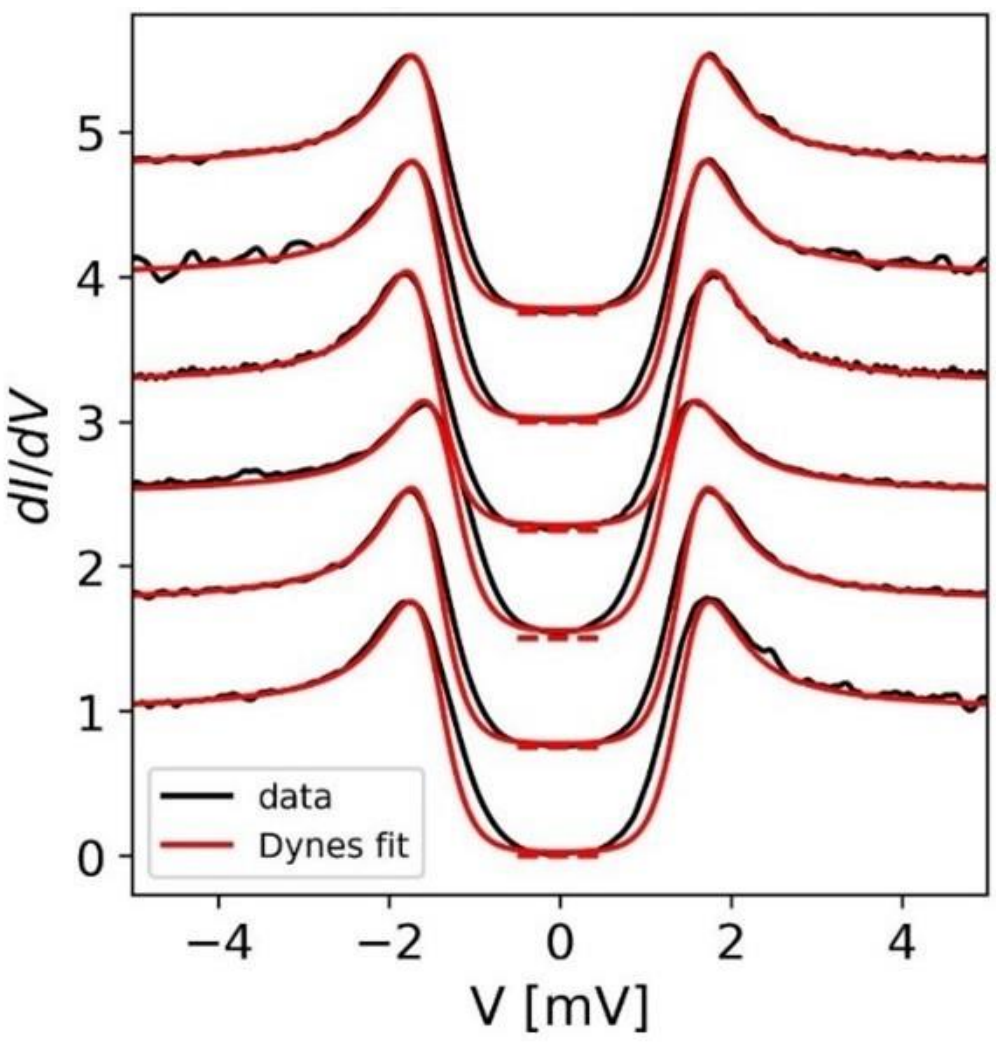

FIG 11: Selected tunneling conductance curves and the corresponding fits from a Nb EP 100 µm + 120 °C 48 h sample. The curves have been vertically shifted for clarity. The red dashed lines indicate the zero-conductance level.

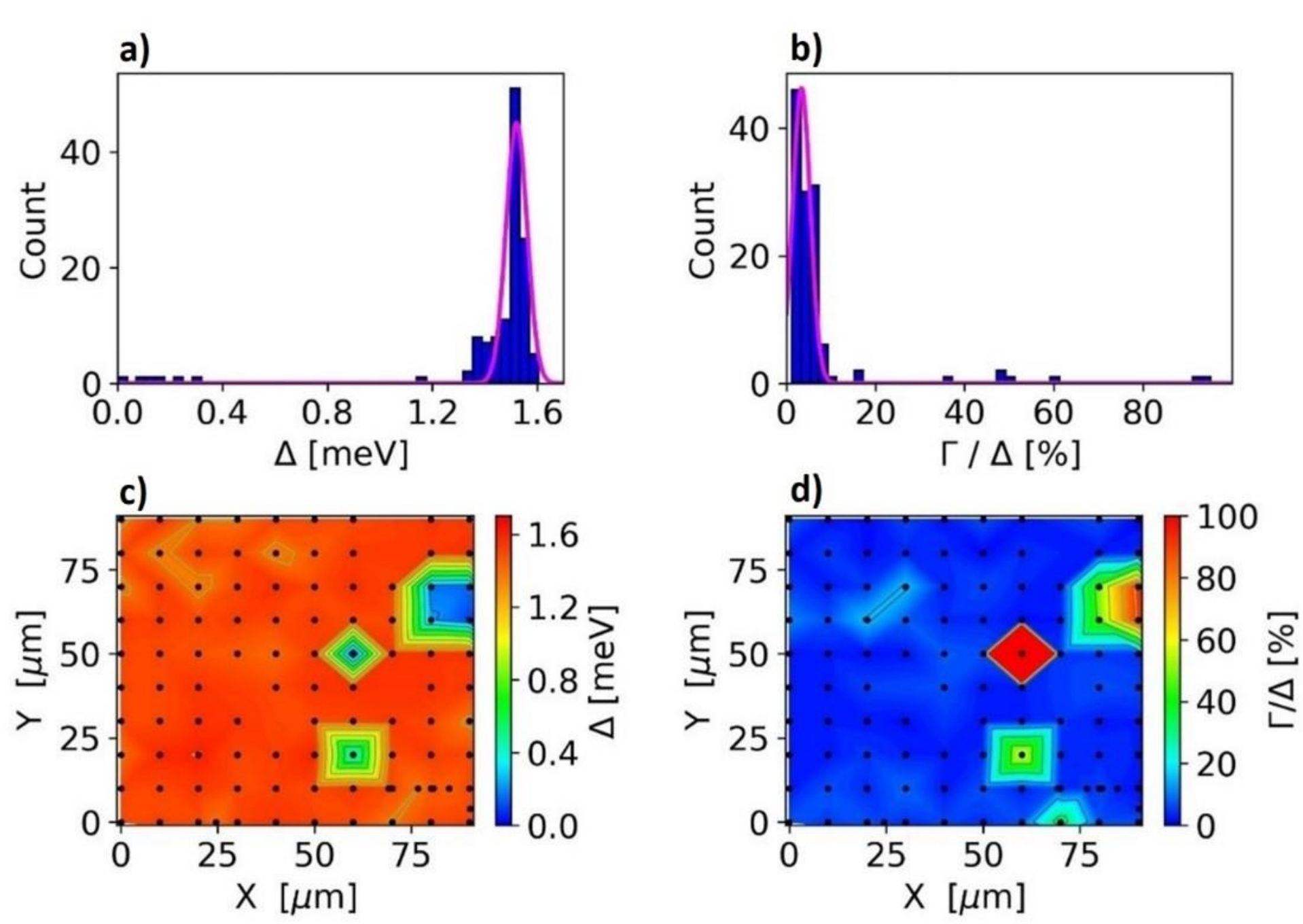

FIG 12: Δ and Γ/Δ statistics and maps obtained from Nb EP 100 µm + 120 °C 48 h sample. In magenta, gaussian (a) and gamma (b) fits. The black dots correspond to the locations of the tunnel junctions.

*Contact author: ivana.curci@cea.fr

These results provide the first direct evidence of the low-temperature baking effect on the surface superconducting properties of electropolished, cavity-grade Nb. The data support a scenario in which low-temperature baking suppresses hydride formation within the RF penetration depth, even under slow-cooling conditions. Our analysis indicates that baking mitigates the formation of regions with reduced superconducting gaps, with the two-step treatment yielding an improvement over the conventional 120 °C bake. We further identify an irreversible influence of cooling after electropolishing, which diminishes the beneficial effect of the subsequent two-step bake, consistent with previous observations [3]. Optimal performance is therefore achieved when cooldown—and, consequently, RF testing—is performed only after completion of all thermal treatments, as is standard practice for SRF cavity testing.

TABLE II. Summary of PCT parameters extracted from Nb samples. The table reports the mean values $\overline{\Delta}$ and $\overline{\Gamma/\Delta}$ extracted from the histogram fits, the averages $\bar{\Delta}$ and $\Gamma/\bar{\Delta}$ computed directly from the data, and the fraction of minigaps.

| sample | | $\overline{\Delta}$ [meV] | $\overline{\Gamma/\Delta}$ [%] | $\bar{\Delta}$ [meV] | $\Gamma/\bar{\Delta}$ [%] | #minigaps |
|---|---|---|---|---|---|---|
| EP 200 μm | before annealing | 1.56 ± 0.04 | 4 ± 3 | 1.24 ± 0.52 | 15 ± 23 | 29 % |
| | after 2stepbake | 1.51 ± 0.03 | 5 ± 2 | 1.43 ± 0.21 | 11 ± 14 | 15 % |
| EP 100 μm | | 1.51 ± 0.08 | 7 ± 6 | 1.45 ± 0.16 | 5.3 ± 3.7 | 14 % |
| EP 100 μm+ 2stepbake | | 1.51 ± 0.02 | 7 ± 2 | 1.51 ± 0.04 | 6.9 ± 2.3 | 1 % |
| EP 100 μm+ 800°C 3h+ EP 10 μm + 2stepbake | | 1.51 ± 0.03 | 1.9 ± 1.5 | 1.51 ± 0.04 | 2.1±1.6 | 1% |
| EP 100 μm+ 120°C 48 h | | 1.50 ± 0.05 | 5 ± 4 | 1.43 ± 0.29 | 8 ± 15 | 6 % |

## IV. DISCUSSION

We now examine the relationship between the local superconducting properties obtained by PCT spectroscopy and the RF response of Nb cavities subjected to equivalent surface treatments. For the two-step baked samples, the mean superconducting gap extracted from the PCT distributions is $\Delta$ = 1.51 meV, in excellent agreement with $\Delta$ = 1.49(3) meV obtained independently from the temperature dependence of the surface resistance, $R_s(1/T)$. By contrast, for the EP samples, Gaussian fits to the PCT gap distributions yield mean values of 1.51–1.56 meV, significantly larger than $\Delta$ = 1.33 meV derived from the $R_s(1/T)$ analysis. This discrepancy can be attributed, at least in part, to the substantial population of junctions exhibiting strongly suppressed gaps, which account for 14–29% of the spectra in the EP samples but contribute only weakly to the Gaussian peak position.
To account for these low-gap regions, we therefore calculate the average gap directly from the complete PCT distributions. This yields $\bar{\Delta}$ = 1.24 ± 0.52 meV and 1.45 ± 0.16 meV for the EP 200 μm and EP 100 μm samples, respectively (Table II). The difference between these values suggests that the superconducting properties of the near-surface region depend sensitively on the amount of material removed by electropolishing. Increased EP has previously been associated with enhanced hydrogen uptake and consequently promote hydride precipitation during cooldown. Importantly, the range $\bar{\Delta}$ = 1.24–1.45 meV encompasses the value $\Delta$ = 1.33 meV extracted from the RF measurements, indicating that the effective superconducting gap probed by the cavity RF response is consistent with a heterogeneous surface containing locally suppressed superconducting regions within a few $\lambda \sim$ 100 nm.

Assuming that the EP 100 μm sample provides a representative model of the EP cavity surface shown in Fig. 1a, we can further examine whether the proximity effect inferred from PCT is quantitatively compatible with the premature onset of the HFQS at $E_{onset} \approx 25$ MV m$^{-1}$. For a TESLA-shaped cavity, this corresponds to a peak surface magnetic field of approximately 106 mT ($H_{MAX}$ (mT) = 4.26 $E_{acc}$ (MV m$^{-1}$) for a TESLA-shaped cavity). Within the proximity-effect model [27], normal-conducting regions embedded in or coupled to superconducting Nb locally reduce the breakdown field, $H_B$, according to

$$H_B \sim \frac{\Phi_0}{6\lambda d} \tag{5}$$

where $\Phi_0$ is the magnetic flux quantum, $d$ is the thickness of the normal region, and $\lambda$ is its effective penetration depth.
Usadel fits to the reduced-gap spectra measured on the EP 100 μm sample yield an average proximity parameter $\alpha \approx 0.15$. Using the assumptions introduced in Eq. (3), together with $\lambda \approx 40$ nm, we obtain a characteristic normal-region thickness of $d \approx 6$ nm, with fitted values spanning approximately between 1–10 nm. This length scale can be compared with that inferred from cavities affected by Q-disease. Such cavities typically exhibit breakdown fields $H_B \approx 10$ mT associated with hydrides approximately 100 nm thick [28–29], corresponding to the empirical scaling $H_B \times d \approx 1000$ mT nm. Applying this scaling to EP cavities with $H_B \approx 106$ mT gives $d \approx 12$ nm. Remarkably, this estimate is comparable to the upper range of thicknesses derived independently from the Usadel analysis. The agreement supports a picture in which nanoscale normal regions, consistent with hydride precipitates, induce proximity-coupled regions of weakened superconductivity that can facilitate premature flux penetration and contribute to the onset of the HFQS.

*Contact author: ivana.curci@cea.fr

The results summarized in Table II further show that the characteristic superconducting gap of the main population remains close to the bulk Nb value for all surface treatments. The principal distinction between treatments is instead the fraction of junctions exhibiting minigaps. Low-temperature baking strongly suppresses this population, with the smallest fraction observed after the two-step treatment. This observation suggests that the beneficial effect of baking does not primarily arise from an enhancement of the intrinsic superconducting gap of Nb, but rather from the elimination or suppression of localized regions with degraded superconductivity.
This connection is illustrated in Fig. 13, where the onset accelerating field of the HFQS, $E_{onset}$—or $E_{MAX}$ for cavities in which no HFQS is observed—is compared with the fraction of minigaps measured by PCT. Representative RF fields for cavities affected by Q-disease and for EP, mild-baked, and two-step baked cavities were obtained from Refs. [5, 21–23, 28–29]. Despite the comparison involving PCT measurements on coupons and RF measurements from different cavities, a clear trend emerges: $E_{onset}$ systematically decreases as the fraction of minigap regions increases. The approximately linear dependence shown in Fig. 13 therefore suggests a direct connection between the prevalence of locally suppressed superconductivity and the field at which losses develop. In this framework, the superior performance of the two-step bake is naturally associated with its ability to nearly eliminate these weak superconducting regions.

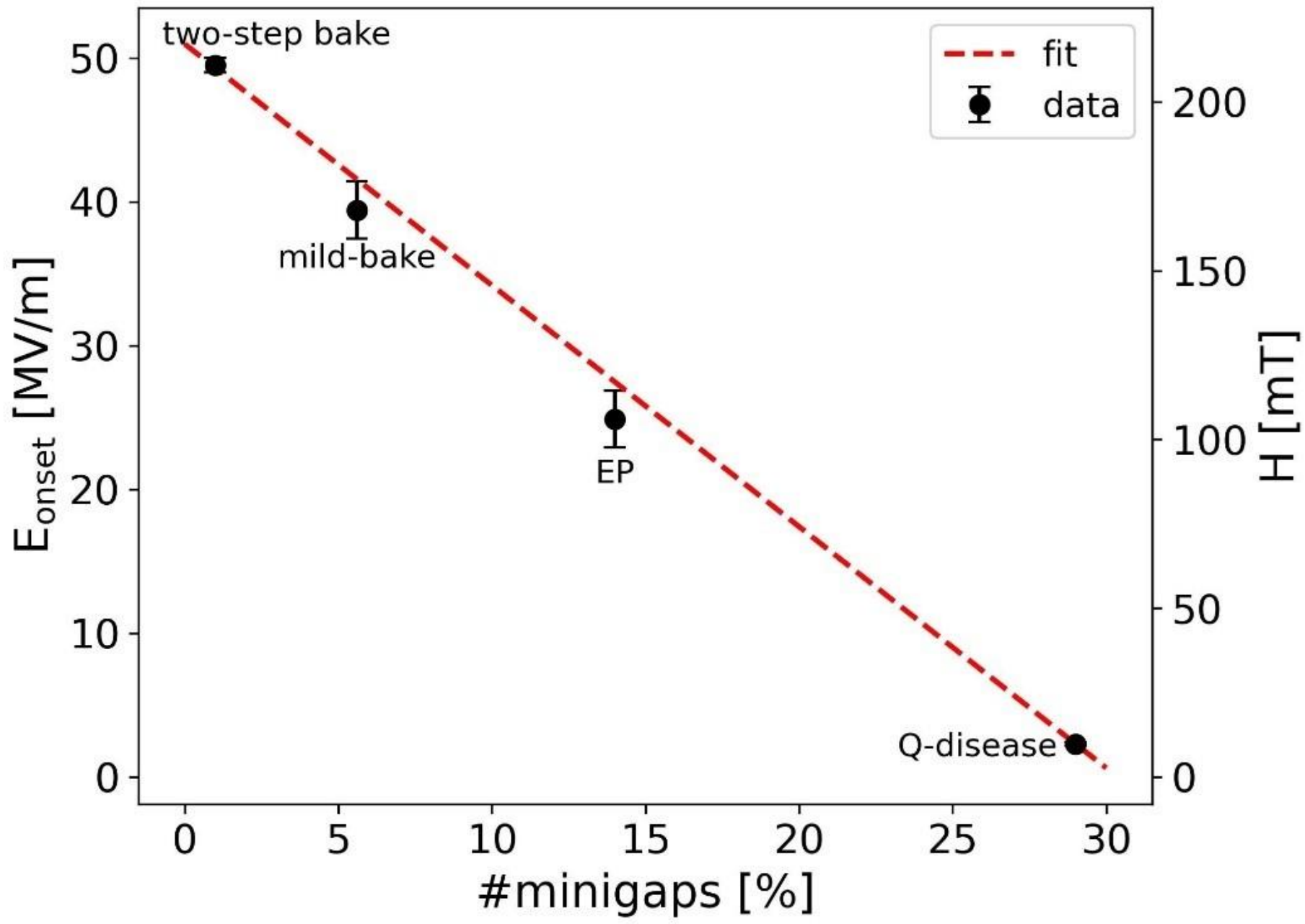

FIG 13: $E_{onset}$ (and the equivalent magnetic field H for a TESLA-shaped cavity) of the HFQS as a function of the fraction of minigaps measured on the samples in Table II.

A second connection between PCT spectroscopy and RF performance emerges from the quasiparticle broadening parameter Γ. Following Ref. [30], the residual resistance associated with finite quasiparticle broadening can be expressed as

$$R_{res} = \frac{\mu_0^2\ \omega^2\ \lambda^3}{2\rho_N}\ \frac{(\Gamma/\Delta)^2}{(1\ +\ \Gamma/\Delta)^2} \quad (6)$$

*Contact author: ivana.curci@cea.fr

where $\mu_0$ is the vacuum permeability, $\rho_N$ is the normal-state resistivity, $\omega = 2\pi f$ is the angular frequency, and $\lambda$ is the London penetration depth. Taking $\rho_N \approx 1$ n$\Omega$ m, $\lambda = 40$ nm, and f = 1.3 GHz for Nb SRF cavities, the experimentally determined residual resistance $R_{res} \approx 5.2$–$7.2$ n$\Omega$ corresponds to $\Gamma/\Delta \approx 4$–$5\%$. The mean $\Gamma/\Delta$ values obtained independently from PCT lie between approximately 5% and 8%, depending on the averaging procedure. Although slightly larger, these values are in good agreement with those inferred from the RF measurements.

Fig. 14 summarizes the dependence of the $R_{res}$ on the $\Gamma/\bar{\Delta}$ measured by PCT. The $R_{res}$ value for the "Q-disease sample" was taken from Ref. [31] for a cavity that stayed 2-3 hrs around 100 K, mimicking the cooling procedure of the PCT sample. The $R_{res}$ value for the EP 100 µm+ 800°C 3h+ EP 10 µm + 2stepbake was measured on a cavity test that followed the same surface treatments at DESY. The fit using Eq. (6) gives a ratio ${\lambda^3}/{2\rho_N}$ = 2 $10^{-5} \pm 1.6$ $10^{-6}$ $m^2\Omega^{-1}$ close to the value of 3.2 $10^{-5}$ obtained from the clean limit values $\rho_N \approx 1$ n$\Omega$ m and $\lambda = 40$ nm (black dashed curve in Fig. 14). More sample statistics is needed to refine these values and confirm the general trend predicted by Eq. (6). This correspondence nonetheless suggests that the finite $\Gamma$ measured by PCT captures microscopic scattering processes that may contribute to the residual RF resistance.

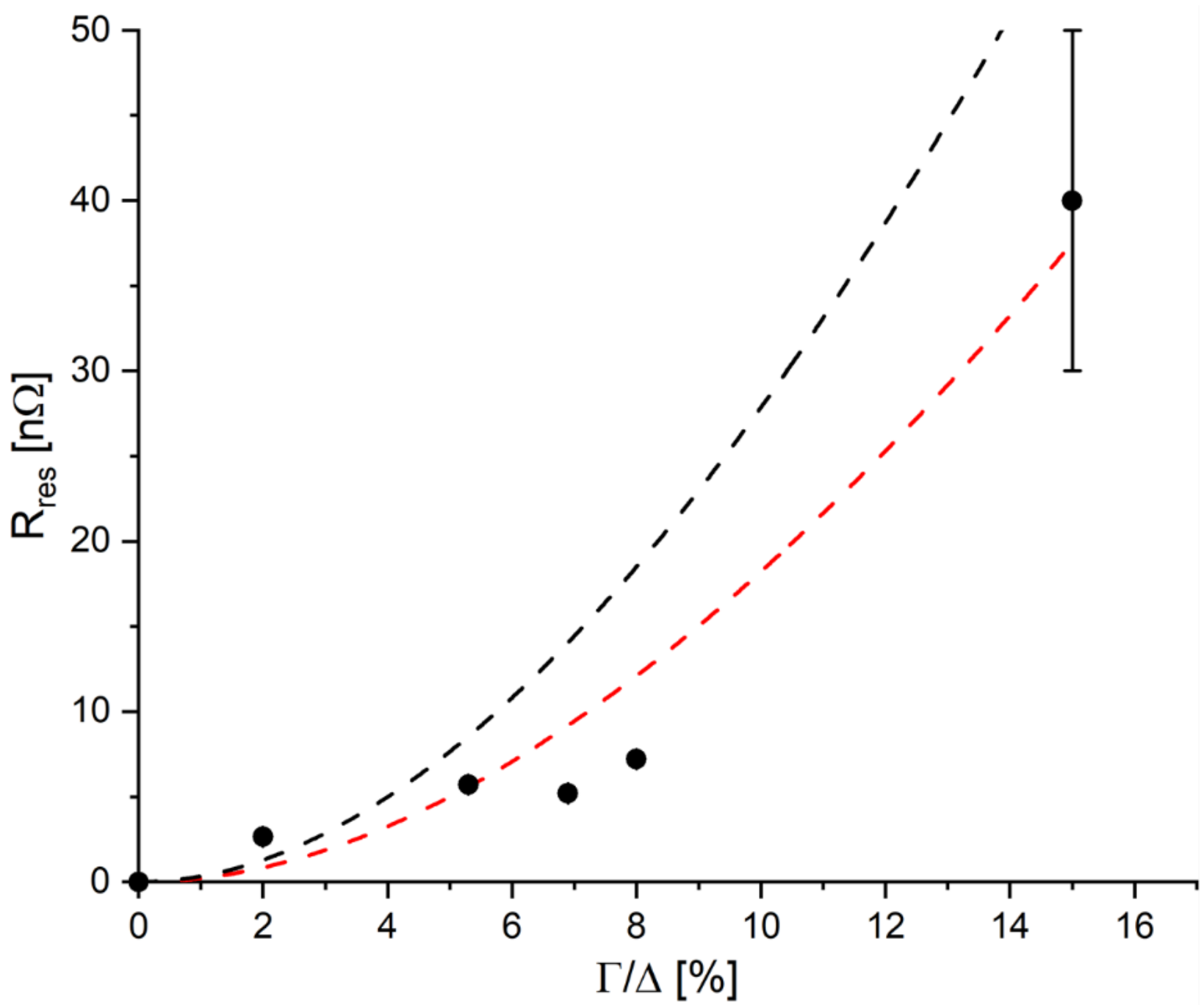

FIG 14: Dependence of $R_{res}$, extracted from the RF measurements, on the $\Gamma/\Delta_{avg}$ obtained from the fits of the tunneling spectroscopy DOS. The dashed red curve are calculated from Eq. (6). The dashed black curve is calculated from the clean limit values $\rho_N \approx 1$ n$\Omega$ m and $\lambda = 40$ nm.

Taken together, these observations reveal two complementary microscopic signatures that can be connected to cavity RF performance. The fraction of minigap regions correlates with the onset of the HFQS, whereas the quasiparticle broadening $\Gamma/\Delta$ is consistent with the magnitude of the residual resistance. The suppression of minigap regions by low-temperature baking, their proximity-effect character, and the nanometer-scale dimensions inferred from the Usadel analysis collectively support a scenario in which nanoscale normal hydrides precipitates locally weaken superconductivity and facilitate premature flux penetration. At the same time, the agreement between $\Gamma/\Delta$ inferred from PCT and RF measurements suggests that inelastic scattering contributes to low-field, temperature independent residual losses. The microscopic origin of the inelastic scattering processes at play is for now unclear and complementary chemical and structural characterization techniques are required. These results demonstrate that local tunneling spectroscopy can provide a microscopic link between near-surface superconducting properties and the macroscopic RF response of Nb cavities.

*Contact author: ivana.curci@cea.fr

## V. CONCLUSION

Point-contact tunneling spectroscopy reveals a clear evolution of the near-surface superconducting properties of cavity-grade Nb with low-temperature baking. Electropolished samples exhibit a significant fraction of regions characterized by suppressed superconducting gaps, whereas these minigap regions are strongly reduced by the 120 °C bake and become nearly absent after the two-step 75 °C/120 °C treatment. Comparison with RF cavity data further reveals a linear relationship between the fraction of minigap regions and the onset field, $E_{onset}$, of the high-field Q-slope (HFQS): treatments producing fewer minigap regions systematically yield higher $E_{onset}$. This correlation identifies locally suppressed superconductivity as a potential microscopic precursor to the HFQS and provides a direct link between local surface properties and macroscopic RF performance.

Conductance spectra measured in minigap regions are well described by the Usadel proximity-effect model, indicating coupling between superconducting Nb and locally non-superconducting regions. The characteristic dimensions inferred from the fits are of the order of ~10 nm, consistent with nanoscale hydride precipitates previously proposed to contribute to the HFQS. Together with the pronounced suppression of minigap regions following low-temperature baking, these observations strongly support a scenario in which baking inhibits hydride formation, thereby reducing proximity-induced weak superconducting regions and extending the accessible RF field range.

Beyond these localized features, the samples exhibit mean quasiparticle broadening parameters $\Gamma/\Delta$ consistent with the residual resistance inferred from the temperature dependence of the surface resistance, $R_s(1/T)$, suggesting that inelastic scattering processes contributing to spectral broadening may also play a role in determining the residual RF losses.

Overall, these results establish a microscopic connection between low-temperature heat surface treatments, local superconducting properties, and the high-field RF response of Nb. They further demonstrate tunneling spectroscopy as a relevant tool for identifying nanoscale regions that limits superconducting resonator (and SRF cavities in particular), performances and provide microscopic insight into the effectiveness of surface treatments.

## ACKNOWLEDGMENTS

This work was supported by Internal CEA Ph.D. funding (CFR) and by the European Union's Horizon 2020 Research and Innovation programme under Grant Agreements No 101004730 and No 101057511 (EURO-LABS).

*Contact author: ivana.curci@cea.fr

*Contact author: ivana.curci@cea.fr

*Contact author: ivana.curci@cea.fr